\documentclass[10pt,conference]{IEEEtran}

\usepackage[utf8]{inputenc}
\usepackage{hyperref}
\usepackage{caption}
\usepackage{subcaption}
\usepackage{tabulary}
\usepackage{textcomp}
\usepackage{graphicx}
\usepackage{xcolor}

\newif\ifarxiv
\arxivtrue

\newcommand{\keyword}[1]{\textit{#1}}
\newcommand{\code}[1]{\textsl{{#1}}}

\newcommand{\apieditor}[0]{API Editor}
\newcommand{\mplt}[0]{Matplotlib}
\newcommand{\skl}[0]{scikit-learn}
\newcommand{\sklv}[0]{0.24.2}

\DeclareRobustCommand{\bird}{%
  \begingroup\normalfont
  \includegraphics[height=\fontcharht\font`\B]{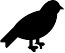}%
  \endgroup
}
\DeclareRobustCommand{\dino}{%
  \begingroup\normalfont
  \includegraphics[height=\fontcharht\font`\B]{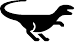}%
  \endgroup
}

\title{Adaptoring: Adapter Generation to Provide an Alternative API for a Library}
\author{\IEEEauthorblockN{Lars Reimann}
\IEEEauthorblockA{Institute of Computer Science III\\
University of Bonn, Germany\\
Email: reimann@cs.uni-bonn.de}
\and
\IEEEauthorblockN{Günter Kniesel-Wünsche}
\IEEEauthorblockA{Institute of Computer Science III\\
University of Bonn, Germany\\
Email: gk@cs.uni-bonn.de}
}

\begin{document}

\ifarxiv
    \IEEEoverridecommandlockouts
    \IEEEpubid{
      \begin{minipage}{\textwidth}
      \noindent\rule[0.5ex]{0.5\textwidth}{0.75pt}\\
      Accepted at the Int. Conf on Software Analysis, Evolution and Reengineering\\
      SANER 2024, March 12–15, 2024, Rovaniemi, Finland\\
      \copyright2024 IEEE. Personal use of this material is permitted.  Permission from IEEE must be obtained for all other uses, in any current or future media, including reprinting/republishing this material for advertising or promotional purposes, creating new collective works, for resale or redistribution to servers or lists, or reuse of any copyrighted component of this work in other works. 
      \end{minipage}
    }
\fi

\maketitle

\ifarxiv
    \IEEEpubidadjcol
\fi

\begin{abstract}
Third-party libraries are a cornerstone of fast application development. To enable efficient use, libraries must provide a well-designed API. An obscure API instead slows down the learning process and can lead to erroneous use.

The usual approach to improve the API of a library is to edit its code directly, either keeping the old API but deprecating it (temporarily increasing the API size) or dropping it (introducing breaking changes). If maintainers are unwilling to make such changes, others need to create a hard fork, which they can refactor. But then it is difficult to incorporate changes to the original library, such as bug fixes or performance improvements.

In this paper, we instead explore the use of the adapter pattern to provide a new API as a new library that calls the original library internally. This allows the new library to leverage all implementation changes to the original library, at no additional cost. We call this approach \keyword{adaptoring}. To make the approach practical, we identify API transformations for which adapter code can be generated automatically, and investigate which transformations can be inferred automatically, based on the documentation and usage patterns of the original library. For cases where automated inference is not possible, we present a tool that lets developers manually specify API transformations. Finally, we consider the issue of migrating the generated adapters if the \emph{original} library introduces breaking changes.
We implemented our approach for Python, demonstrating its effectiveness to quickly provide an alternative API even for large libraries.
\end{abstract}

\begin{IEEEkeywords}
API design, API evolution, Adapter pattern, Wrapper library, Refactoring, Adaptoring
\end{IEEEkeywords}




\section{The API Evolution Challenge}
\label{sec:introduction}

\begin{figure*}[t]
    \centering
    \begin{subfigure}[t]{0.18\textwidth}  
        \centering 
        \includegraphics[height=1.2cm]{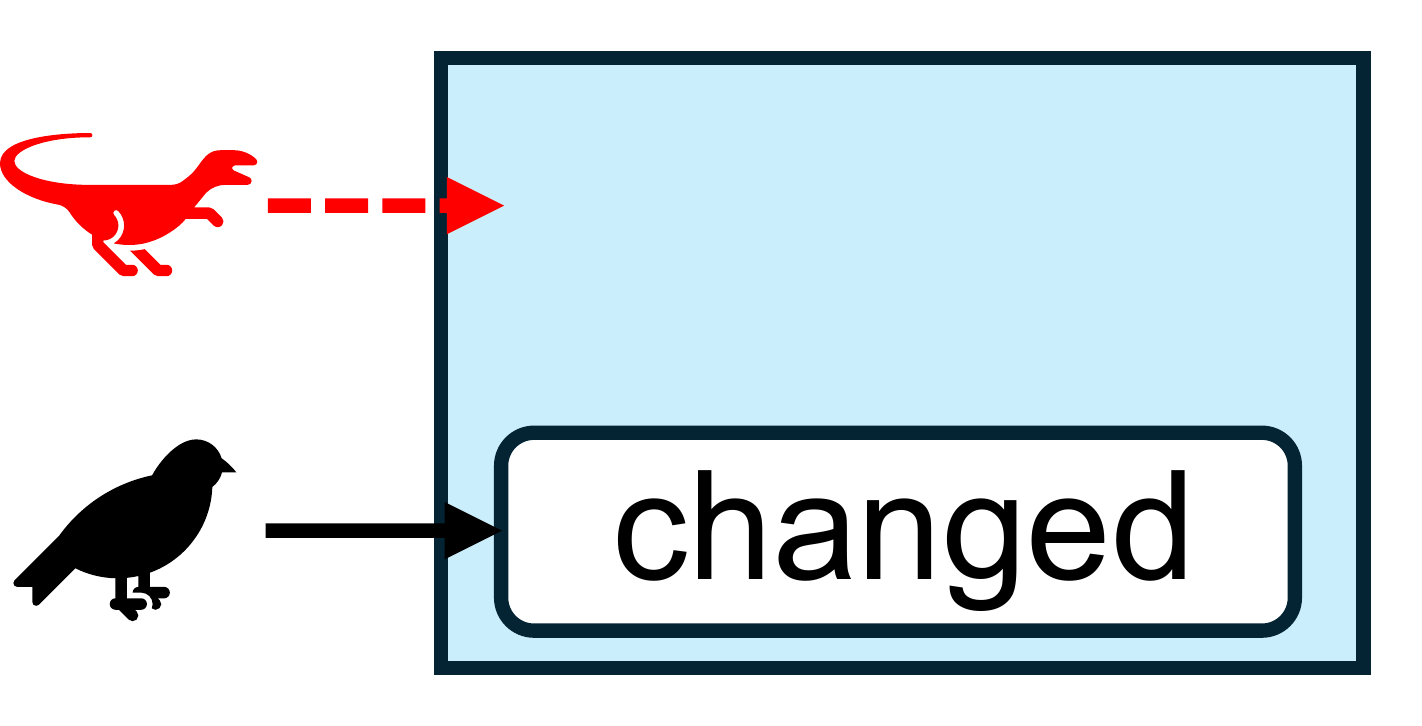}
        \caption{Refactoring, drop the original; breaking clients}
        \label{fig:approaches-for-api-transformation/refactoring-remove-original}
    \end{subfigure}
    \hspace{0.65em}
    \begin{subfigure}[t]{0.18\textwidth}
        \centering 
        \includegraphics[height=1.2cm]{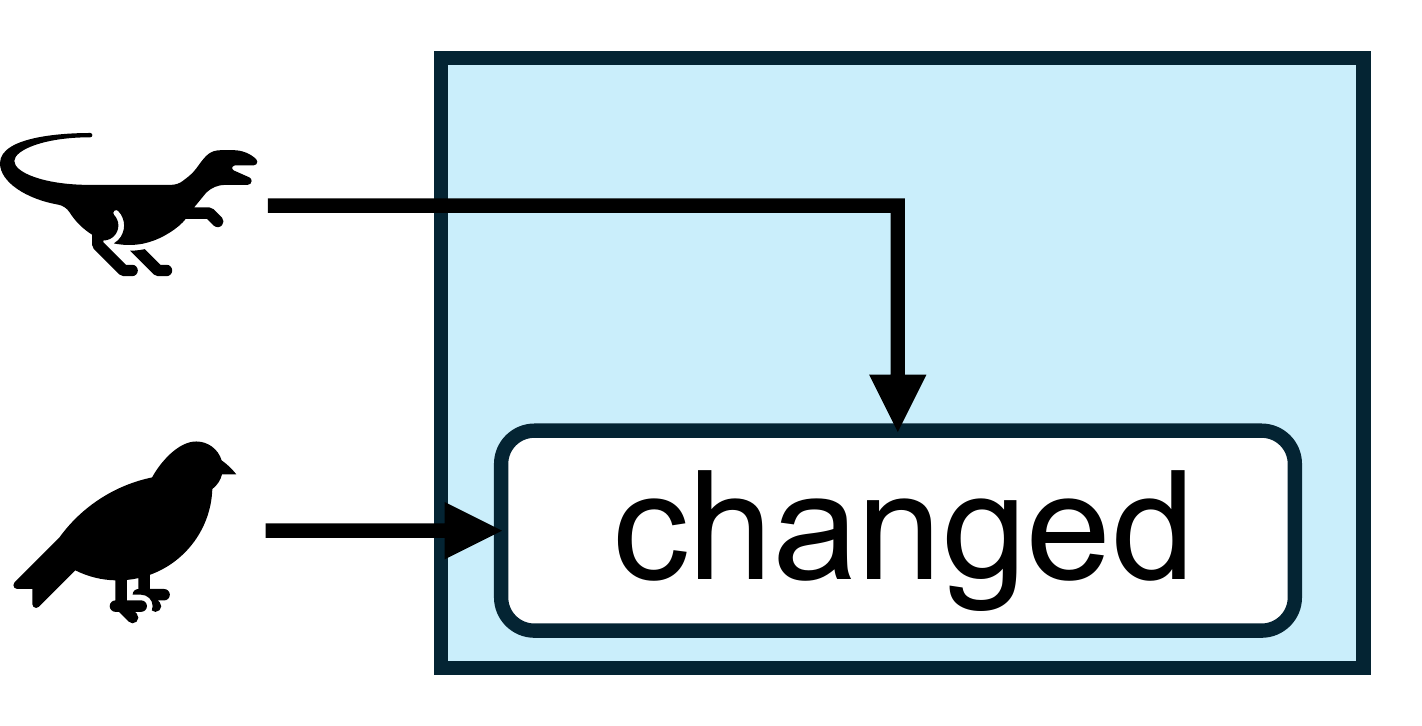}
        \caption{Refactoring, drop the original, migrate clients}    
        \label{fig:approaches-for-api-transformation/refactoring-migrate}
    \end{subfigure}
    \hspace{0.65em}
    \begin{subfigure}[t]{0.18\textwidth}
        \centering
        \includegraphics[height=1.2cm]{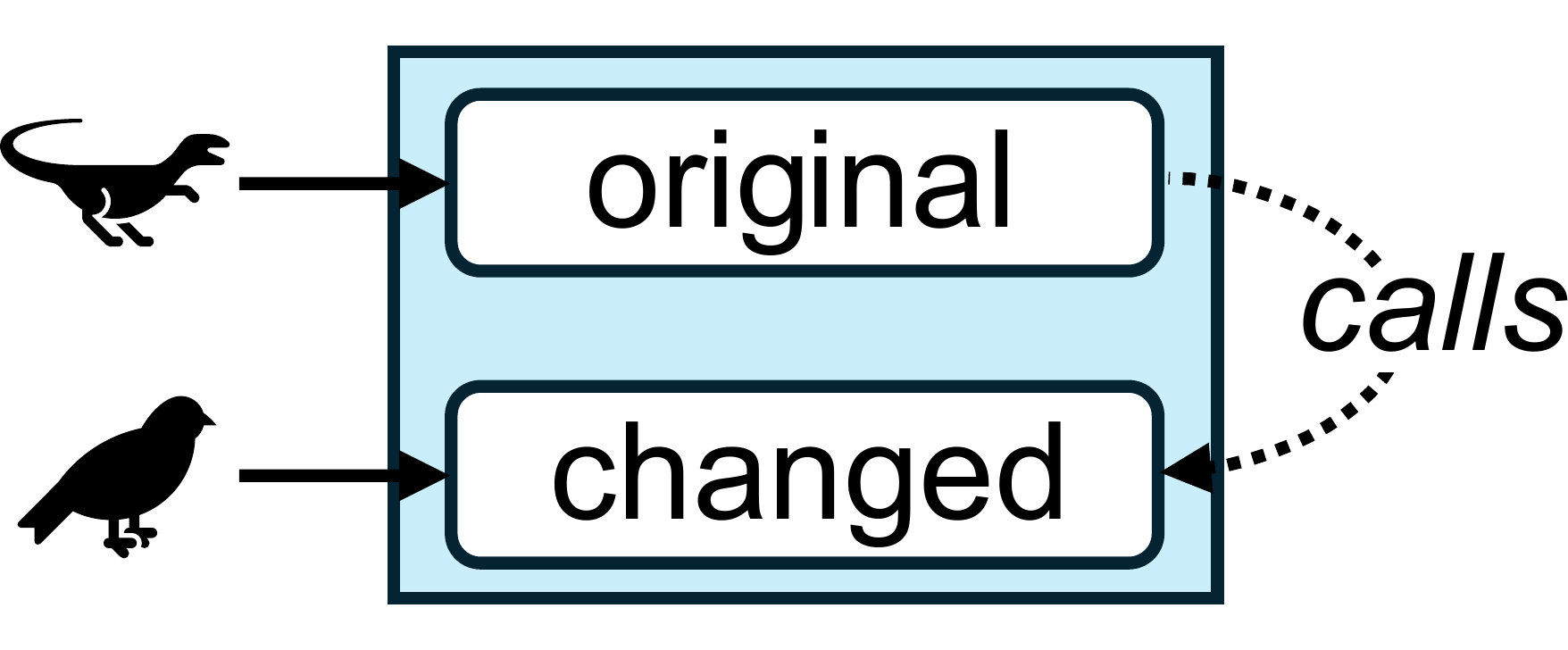}
        \caption{Refactoring, keep the original but deprecate it}  
        \label{fig:approaches-for-api-transformation/refactoring-keep-original}
    \end{subfigure}
    \hspace{0.65em}
    \begin{subfigure}[t]{0.17\textwidth}   
        \centering 
        \includegraphics[height=1.2cm]{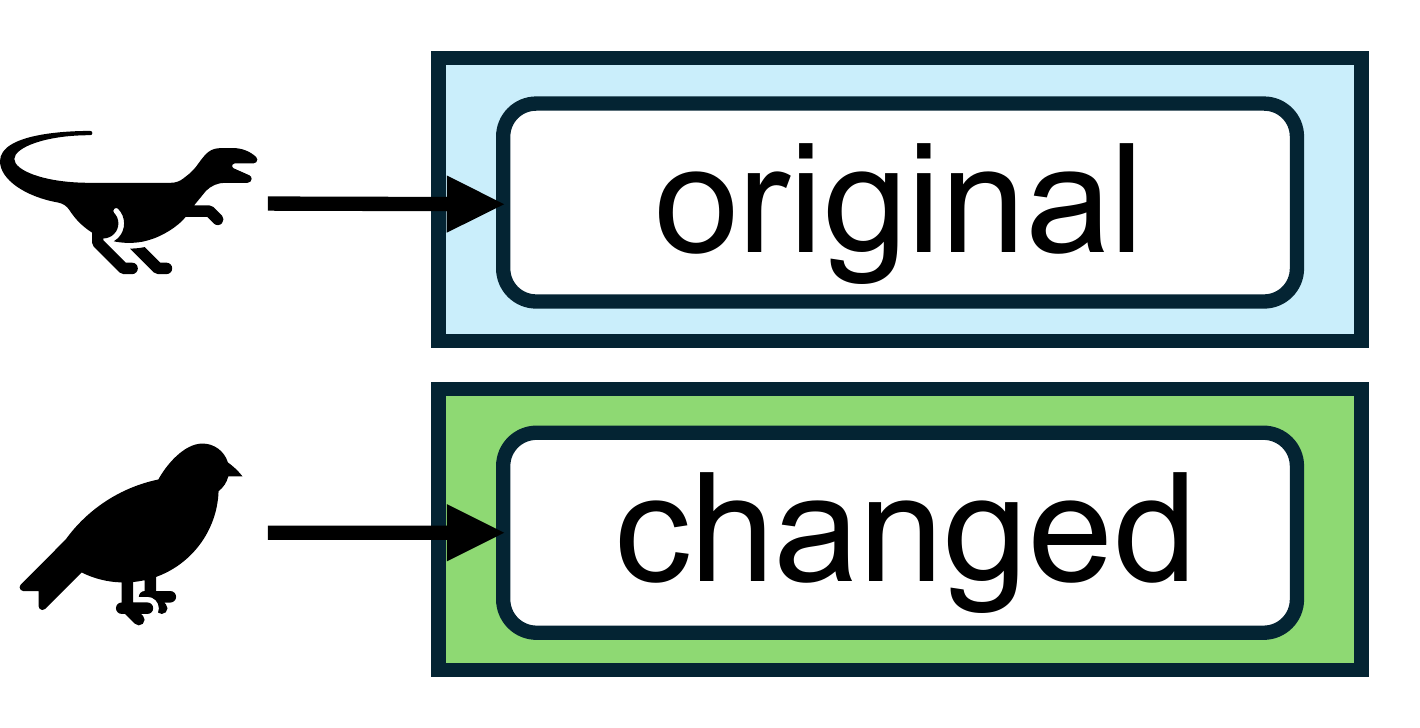}
        \caption{Hard-forking (two decoupled code bases)}   
        \label{fig:approaches-for-api-transformation/hard-fork}
    \end{subfigure}
    \hspace{0.65em}
    \begin{subfigure}[t]{0.19\textwidth}   
        \centering 
        \includegraphics[height=1.2cm]{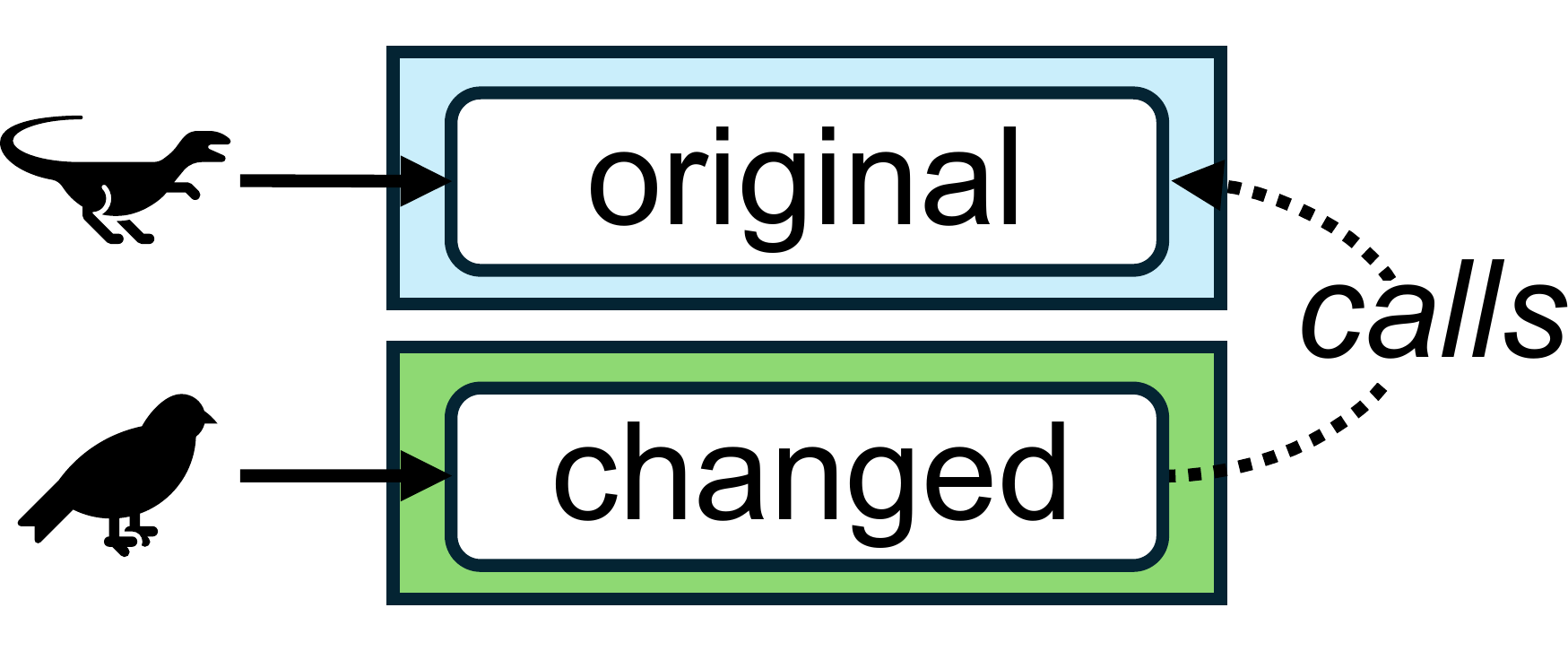}
        \caption{Adaptoring (two APIs sharing original code base)}   
        \label{fig:approaches-for-api-transformation/adaptoring}
    \end{subfigure}
    \caption{Discussed approaches for creating a new API for a library. Blue boxes represent the original library and green boxes the new one. Original users (\dino{}) interact with the original function, unless they migrate their own code, while new users (\bird{}) directly use the changed function. Adaptoring builds a new API based on the unmodified, original one.} 
    \label{fig:approaches-for-api-transformation}
\end{figure*}

Using third-party libraries allows application developers to focus on their core business logic. They no longer have to reinvent the wheel, to solve recurrent issues like the implementation of an encryption algorithm, but can instead rely on an existing and well-tested solution. To interact with a library, application developers need to understand its application programming interface (API), which places a particular emphasis on the learnability and usability of said API \cite{desouzaAutomaticEvaluationAPI2009,WangGodfrey:MSR2013,ZibranEtAlWCRE2011}. 
Often, however, the API is complex and suffers from design issues that crop up in software developed by many contributors over many years \cite{Robillard:IEEESOftware2009,Robillard:ESE2011}.
The maintainers of the affected library can remedy these problems in different ways, summarized in Fig. \ref{fig:approaches-for-api-transformation} and discussed below. 

\section{State of the Art}
\label{sec:state-of-the-art}

\ifarxiv
    \IEEEpubidadjcol
\fi

\paragraph*{API refactoring} 
\label{sec:soa:refactoring} 
The straightforward way of improving an API is to change the original library directly. The systematic way of direct change is via refactorings. In the second edition of his book, Fowler \cite{refactoring} devotes an entire chapter to API refactoring. 
Dig and Johnson \cite{digRoleRefactoringsAPI2005} analyzed API changes and found that 80\% of them were basic refactorings. 
Koçi et al. \cite{kociClassificationChangesAPI2019} similarly classified API changes. 
IDEs like Eclipse\footnote{\url{https://www.eclipse.org/ide/}} or IntelliJ IDEA\footnote{\url{https://www.jetbrains.com/idea/}} offer various automated refactorings that can be used to alter an API. 
For broader coverage of the vast refactoring literature, we refer to prior surveys \cite{mensSurveySoftwareRefactoring2004, abebeTrendsOpportunitiesChallenges2014, baqaisAutomaticSoftwareRefactoring2020}.

\paragraph*{Automated migration of client code} 
After changing an API, client code might have to be adjusted. To avoid tedious and error-prone manual adjustment, 
Perkins \cite{perkinsAutomaticallyGeneratingRefactorings2005} suggests that API developers provide code to replace calls to deprecated methods. 
Henkel and Diwan \cite{henkelCatchUpCapturingReplaying2005} capture refactorings during development of the API, which can then be replayed for the client code. 
Similarly, Hora et al. \cite{horaAPIEvolutionMinerKeepingAPI2014} monitor changes to the API of a library as its source code is altered. 
Xing and Stroulia \cite{xingAPIEvolutionSupportDiffCatchUp2007} detect changes automatically without additional work from the API developers and derive plausible replacements for client code.
In a case study on an Android API, Lamothe and Shange \cite{Lamothe2018ExploringTU} explore the use of automated API migration techniques and report that  determining the types of new parameters is still a challenge for the reviewed approaches.  

\paragraph*{Keep deprecated API}
Breaking changes and migration issues can be deferred, by temporarily keeping the old API as methods that forward calls to the new API. Typically, these forwarding methods are deprecated. The deprecation message lets users schedule migration and can include migration guidance. However, Kao et al. \cite{kao-api-deprecation} point out that ``API deprecation entails significant cost to API developers'' including ``code bloat [...] and serious threats to security, performance, or code quality [and] the cost of migration guide provision.'' In addition, \emph{new} clients need to deal with an unnecessarily large API, which impedes learning and correct use.

\paragraph*{Hard forking} 
If library maintainers are not willing to change the API in any of the ways above, other developers need to create a hard fork of the library that they can then refactor. This, however, decouples the fork from the original library, making the integration of future changes difficult. 


%

\paragraph*{Adapter generation} 

A different way to provide a new API for a library
is proposed by Jugel \cite{jugelGeneratingSmartWrapper2010}: It allows deleting and renaming API elements of a Java library in a UML editor. Afterward, a transformation is applied to create a DSL-like API using the builder pattern \cite{gof}. Finally, code gets generated to implement the API using the adapter pattern \cite{gof}. 


\section{A New Approach: Adaptoring}
\label{sec:adaptoring-approach}

Fig. \ref{fig:approaches-for-api-transformation} summarizes the different approaches for creating an alternative API that we discussed.
Among the reviewed approaches, the use of the adapter design pattern \cite{gof} is the only way to provide a new API for a library without breaking clients (Fig. \ref{fig:approaches-for-api-transformation/refactoring-remove-original}), raising migration issues (Fig. \ref{fig:approaches-for-api-transformation/refactoring-migrate}), bloating library code by deprecation (Fig. \ref{fig:approaches-for-api-transformation/refactoring-keep-original}), or completely duplicating library code (Fig. \ref{fig:approaches-for-api-transformation/hard-fork}). 

However, 
we only found a single prior work that suggests an adapter-based approach \cite{jugelGeneratingSmartWrapper2010}. Unfortunately, it suffers from several limitations:

\begin{itemize}
    \item No implementation of the approach is publicly available.
    \item The only supported API transformations are renamings and deletions, although other API transformations can be realized by adapters (e.g. moving an API element).
    \item  To quickly create a wrapper library for a large library, automation is vital. However, \cite{jugelGeneratingSmartWrapper2010} does not automatically infer potential transformations.
    \item The usability of the editor for manual API transformation is lacking. This was criticized by the author himself.
    \item There is no support for the evolution of the adapter layer after breaking changes to the original API.
\end{itemize}

We think that it is worthwhile to explore ways to overcome the above-mentioned limitations because an adapter-based approach has clear advantages, for library maintainers as well as application developers:

\emph{Library maintainers} can use an adapter layer to provide an alternative API that is clearly separated from the original API. New clients can directly use the self-contained alternative API, while existing clients have time to migrate from the original API to the alternative one, perhaps aided by automated migration approaches.
%
They can additionally support separate APIs for different user profiles: For example, the parts of an API intended to enable extensions might not be needed by application programmers, who are interested in using the library for their business logic. Likewise, novices would benefit from a smaller API that provides only commonly needed API elements. This accelerates development and can prevent misuses resulting from misunderstanding the function of overwhelmingly many API elements \cite{myersImprovingAPIUsability2016, schellerInfluencingFactorsUsability2012}.

\emph{Application developers} can use an adapter layer as an alternative to hard forking, if the maintainers do not want to create it themselves: Since the original library is called by the adapters, any implementation changes, like bug fixes or performance improvements, can be consumed without changes to the wrapper library. Only changes to the original API, require follow-up changes to the related adapters. The adapter layer can be made available to other developers, for example for in-house development, so the effort to create it initially need not be repeated.

In the remainder of the paper, we will elaborate \emph{adaptoring}, as a refinement of the adapter-based approach that overcomes some of the limitations of \cite{jugelGeneratingSmartWrapper2010}. 
Adaptoring creates a wrapper library to provide an alternative, more usable API for an existing library (Fig. \ref{fig:approaches-for-api-transformation/adaptoring}).
This leads us to our main research questions:

\begin{itemize}
    \item RQ 1: Which API transformations can be supported? (Sec. \ref{sec:adaptoring-approach/adaptorable-api-transformations})
    \item RQ 2: How can API transformations be inferred automatically? (Sec. \ref{sec:inferring-api-transformations})
    \item RQ 3: How can the manual specification of API transformations be eased? (Sec. \ref{sec:api-editing})
    \item RQ 4: How can the adapter code be evolved if the original library changes? (Sec. \ref{sec:evolution-of-original-library})
\end{itemize}


\section{Adaptorable API transformations}
\label{sec:adaptoring-approach/adaptorable-api-transformations}
We call an API transformation that can be implemented by using the adapter pattern \cite{gof} \keyword{adaptorable}. 
Via an adaptorable transformation, an original API element is 
\begin{itemize}
    \item left out, 
    \item moved to another location,
    \item modified regarding its name, parameters, parameter types, preconditions, or language-specific attributes (such as the Python-specific optionality of parameters)
\end{itemize}
or a new API element is created that provides the desired API by invoking functions of the original library.
As additional code besides the calls to the original library, we only allow the validation of preconditions and code that is syntactically required to achieve the desired new API, such as a new class that bundles several functions. No new behavior may be added.


%
%
We use Fowler's refactorings \cite{refactoring}, constrained to refactorings that actually change the API, as our base catalog of systematic API transformations. However, not all the API refactorings can be handled by adaptoring. Particularly, we can only add new API elements whose implementation can be based on the original API. This excludes refactorings like \code{Extract Function}\footnote{We assume the extracted function becomes part of the public API.} or \code{Replace Conditional with Polymorphism}. All API transformations that can be achieved by adaptoring can also be achieved by refactoring, since we could also add the adapter to the original library, inline the calls, and replace the original function with it. Thus, the adaptorable API transformations are a subset of the API transformations that refactoring can accomplish. This still leaves a substantial set of API-transforming refactorings that \emph{can} be realized by adaptoring (RQ 1): 

\begin{itemize}
    \item API elements can be removed by simply not creating adapters for them. This way, useless or redundant elements can be eliminated, which decreases the size of the API and helps the user focus on the relevant parts. The removal also accelerates the redesign, since these API elements need not be changed further.
    \item Parameter lists of callables can be reordered, or language-specific attributes like optionality and the default value of parameters can be changed.
    %
    \item Preconditions can be enforced, e.g. by introducing assertions\footnote{Since additional exceptions are thrown when preconditions are violated, this is an API change.} or using narrower types for function parameters. This makes the API more robust. 
    \item API elements can be renamed to a more understandable name. This also helps their discoverability.
    \item API elements can be moved so that they are close to related API elements and can be found more easily.
    \item API elements can be pushed to subclasses if they are only relevant for some of them.
    \item API elements can be repackaged (e.g. \code{Introduce Parameter Object} or \code{Replace Function with Command}).
\end{itemize}

\paragraph*{Implementation for Python}

To test our approach, we implemented a prototype that can apply various API transformations to Python libraries and generate adapters for them (Sec. \ref{sec:adaptoring-approach/adapter-generation}). 
%
%
%
So far, our prototype implements the adaptorable API transformations shown in Table \ref{tab:implemented-transformations}. For many of them, we implemented automated suggestions (Sec. \ref{sec:inferring-api-transformations}), as indicated by the ``Inferred'' column. The ``Not yet'' for the \code{Move} transformation means that approaches to infer moves \cite{MoveRecommendationsUsingPathRepresentation, TERRA201819, 10.1145/3275245.3275247, 6684534, moveRecommendationsUsindDependencySets} exist and could be integrated. We will discuss other features of the prototype in the next sections.

\begin{table}[ht]
    \centering
    \caption{API Transformations Implemented for Python}
    \begin{tabulary}{.503\textwidth}{|l|L|l|} 
        \hline
        \textbf{Transformation} &
        \textbf{Semantics} &
        \textbf{Inferred} \\
        \hline\hline
        \multicolumn{3}{|c|}{\textbf{Deletions (Sec. \ref{sec:inferring-deletions})}} \\
        \hline
        \code{Remove} &
        Remove class or function &
        Yes \\
        \hline
        \code{Replace with constant} &
        Remove parameter from new API and internally replace references to it by a constant &
        Yes \\
        \hline\hline
        \multicolumn{3}{|c|}{\textbf{Optionality of Parameters (Sec. \ref{sec:inferring-optionality-of-parameters})}} \\
        \hline
        \code{Make optional} &
        Make parameter optional and specify its default value & 
        Yes \\
        \hline
        \code{Make required} &
        Make parameter required &
        Yes \\
        \hline\hline
        \multicolumn{3}{|c|}{\textbf{Enforcement of Preconditions (Sec. \ref{sec:inferring-enforcement-preconditions})}} \\
        \hline
        \code{Add bounds check} &
        Ensure parameter value is within an interval & 
        Yes \\
        \hline
        \code{Replace with enum} &
        If a parameter can take only a finite set of strings, encode these values in an enum and replace the string parameter with a parameter that accepts instances of this enum & 
        Yes \\
        \hline\hline
        \multicolumn{3}{|c|}{\textbf{Other Design Changes}} \\
        \hline
        \code{Rename} &
        Rename declaration & 
        No \\
        \hline
        \code{Move} &
        Move declaration & 
        Not yet \\
        \hline
        \code{Group} &
        Group parameters into parameter object \cite{refactoring} & 
        No \\
        \hline
    \end{tabulary}
    \label{tab:implemented-transformations}
\end{table}

\section{Adapter Generation}
\label{sec:adaptoring-approach/adapter-generation}


Creating adapters \cite{gof} for a large library manually is tedious, error-prone, and time-consuming and should, therefore, be automated.
%
Given a list of adaptorable transformations, we use the following five-step process to generate adapters:

\begin{enumerate}
    \item Collect a list of all API elements of the original library, including classes, functions and their parameters.
    \item Create an abstract syntax tree (AST) that represents trivial wrappers around each function of the original library. Trivial wrappers apply no changes yet. Each contains just a call to the function with the exactly same signature from the original library. 
    Compared to refactoring \cite{refactoring}, the forwarding direction is inverted: When adaptoring, the new function calls the old one, while when refactoring, the old function can be kept to forward to the new one to prevent breaking changes.
    \item Adjust this AST by applying the specified API transformations one by one. This may delete the wrapper altogether, change its signature (name, parameters), or change its implementation. Implementation changes include changes to the arguments of the call to the original library and insertion of argument checking operations.
    \item Post-process the resulting AST to ensure correctness of the output created in the next step. In this step, we can enforce language-specific constraints, which may be violated after applying the API transformations. 
    \item Generate code by visiting the nodes in the AST and serializing them. This creates code for all adapters that were not removed, including unchanged ones, so no interaction with the original library is necessary for users of the new one.
\end{enumerate}

 \paragraph*{Implementation for Python} In our Python prototype, the adapter generator implements the algorithm outlined above. To collect API elements in Step 1, we use astroid\footnote{\url{https://github.com/pylint-dev/astroid}}. So far, the prototype only deals with classes, functions, and parameters (differentiating between required and optional parameters), but can be extended easily to other kinds of API elements such as fields. In Step 4 (post-processing), we reorder parameters to ensure that all optional parameters are listed after all required ones, which is mandated by Python. Finally, in Step 5 we also include the docstrings of the corresponding API elements of the original library in the generated code, without the sections referring to removed API elements.

\section{Inferring API Transformations}
\label{sec:inferring-api-transformations}

Adapter code generation can be automated as described above, given a list of adaptorable API transformations (Sec. \ref{sec:adaptoring-approach/adapter-generation}). This leaves the specification of such a list as the remaining problem. In this section, we discuss the automated inference (RQ 2) of deletions (Sec. \ref{sec:inferring-deletions}), parameter optionality (Sec. \ref{sec:inferring-optionality-of-parameters}), and precondition checks (Sec. \ref{sec:inferring-enforcement-preconditions}). In Sec. \ref{sec:api-editing} we discuss manual specification of transformations. 
We limit our discussion to classes, functions, and parameters (differentiating between required and optional parameters), but the notion can be extended to other kinds of API elements such as fields.

\subsection{Inferring Deletions}
\label{sec:inferring-deletions}

\paragraph*{Unused elements} Our first focus is the reduction of an API's size, to limit the amount of work necessary in subsequent steps. 
Our analysis counts \keyword{usages} of API elements in an (ideally large) set of client programs:
\begin{itemize}
    \item A usage of a class $c$ is a call to a method of $c$.
    \item A usage of a function $f$ is a call of $f$.
    \item A usage of a parameter $p$ is the explicit setting of the parameter $p$ when the containing function is called.
\end{itemize}
Counting usages of an API element across all analyzed client programs yields the \keyword{usage count} of that element. We call elements with usage count $0$ \keyword{unused} and all others \keyword{used}. Unused elements can be removed without breaking any of the analyzed client programs\footnote{Other client programs that we did not analyze might still use the API element, so element deletion must still be considered a breaking change.}.

\paragraph*{Useless elements}
For \emph{new} clients, we can disregard backward compatibility, enabling further elimination of used but useless parameters. 
We start identification of useless parameters by counting how often a parameter $p$ is set to a particular value $v$. 
We do not distinguish whether $p$ is set to $v$ explicitly, or whether $v$ is set implicitly as the default value of an optional parameter if the language permits.
Based on the parameter value counts, we identify the most common value for each parameter and define the \keyword{usefulness} of a parameter $p$, as the number of times $p$ has been set, explicitly or implicitly, to a value \emph{different from} the most common one. 
%
We call 
a parameter \keyword{useless} if it is always set to the same value. A \keyword{useful} parameter is set to at least two different values. Each useless parameter can be eliminated from the respective function and set to its unique constant value internally, in the implementation of the function. This simplifies the API for new clients, but is a breaking change for existing client programs that set the parameter explicitly.

\paragraph*{Rarely used and almost useless elements}
By eliminating useless parameters, we create an API that will serve new clients well but will not be backward compatible. Once we have crossed this line, further reductions can be achieved for new users.
To this end, we generalize the notion of unused and useless API elements: Given a threshold $T \geq 0$, we call an API element \keyword{rarely used} if its usage count is less than $T$. Likewise, we call an element \keyword{almost useless} if its usefulness is less than $T$. 
In this context, we define the usefulness of classes and functions to be equal to their usage count.
%
%
To achieve an even stronger size reduction, almost useless elements can be additionally removed from the API. Alternatively, we could define a basic API version for novices and an expert version of an API, as suggested by \cite{xuHeyYouHave2015a}, using different thresholds. The terms unused and useless describe the special case $T = 1$. Setting $T = 0$ yields the original API, without any changes.




\paragraph*{Implementation for Python} 
In our prototype, we analyze client code statically using \code{astroid} to build an AST, which is then traversed to find and count calls. \code{astroid} is also used to derive call targets, where possible (this is non-trivial, in Python). Afterward, class, function, and parameter usages are counted as described above. 
To compute value counts for a parameter, all occurrences of non-literal values are differentiated. This prevents wrongly marking a parameter as useless (see Sec. \ref{sec:inferring-deletions}) if it is always passed a reference to the same local variable \code{x}, which could refer to many different values, particularly in different client programs. No partial evaluation is implemented currently, so the expressions \code{2} and \code{1 + 1} are seen as different values.
After counting, API elements with a usefulness below the set threshold are marked for deletion (see Sec. \ref{sec:inferring-deletions}). The entire analysis is available as part of the
\code{library-analyzer}
Python package\footnote{\url{https://github.com/Safe-DS/Library-Analyzer}}.

Derivation of the call target is challenging in Python due to its dynamic features. Better results could be obtained by dynamic analysis. For our prototype, we chose static analysis primarily for security reasons, so that users do not run thousands of client programs mined from the internet on their machine. Moreover, in our application to data science, client programs often take minutes to several hours to run, which makes dynamic analysis prohibitive. 
%


\subsection{Inferring Optionality of Parameters}
\label{sec:inferring-optionality-of-parameters}

Python and other programming languages let developers mark parameters as required or optional. The choice between making a parameter required or optional is a trade-off between correctness and usability. More optional parameters mean less effort for programmers, who do not need to explicitly set a value. More required parameters mean less risk that developers might not notice that values other than the default one would be more appropriate. Therefore, defining default values for parameters needs thoughtful consideration and should only be done if there is a clear most common value, in the first place. For example, a boolean parameter that is almost equally often set to \code{True} and \code{False} should not be optional but required, to force developers to choose consciously.


%
Our approach is to make a parameter optional only if we have sufficient empirical evidence that it has a clear most common value. For this, we do a \emph{null-hypothesis significance test} \cite[p.~923]{gowers08} 
by looking only at the two most common values of a parameter $p$, let's call them $v_1$ (most common) and $v_2$ (second most common). Let $uc_1$ and $uc_2$ respectively denote how often $v_1$ and $v_2$ are used (explicitly or implicitly), and $n := uc_1 + uc_2$. We then assume the null hypothesis $H_0$ that in the $n$ cases where either value was used, users chose $v_1$ and $v_2$ equally often. If we can reject $H_0$, we make the parameter optional with $v_1$ as its default value. Otherwise, we make it required. The probability distribution underlying $H_0$ is a binomial distribution \cite[p.~263]{gowers08} with a success chance of 0.5, and we reject $H_0$ if its p-value \cite{enwiki:p-value} is less than or equal to a significance level $\alpha$ in a two-sided test\footnote{Due to the symmetry of the binomial distribution, this is equivalent to a one-sided test with a significance level of $\frac{\alpha}{2}$.}:
$$
2\sum_{i = uc_1}^{n}{{n \choose i} 0.5^{n}} \leq \alpha
$$

\noindent
The significance level $\alpha$ can be tuned to produce a suitable result. A lower $\alpha$ produces more required parameters, while a higher $\alpha$ produces more optional parameters. The formula is designed so that parameters in rarely used functions (as long as we don't have more data about actual usages) are more likely to be made required, opting for safety in this case. It prevents breaking changes when the default value is changed or removed. Usability is not affected noticeably, since these functions are rarely used, anyway.

\paragraph*{Implementation for Python} 
The value count analysis of the \code{library-analyzer}
package that we used to delete (almost) useless parameters (Sec. \ref{sec:inferring-deletions}) already provides the data we need for this inference step. The API redesigner can set the significance level, $\alpha$ and the suggestions for the optionality of parameters are created based on the above formula. The limitations of static analysis also apply here: In particular, we never change the optionality of a parameter if the most common or second most common values are non-literals (e.g. variable references), since we do not know their actual value.

\subsection{Inferring Enforcement of Preconditions}
\label{sec:inferring-enforcement-preconditions}

Preconditions can also inspire API transformations, as shown in the following non-exhaustive list. All can be enforced by the addition of runtime checks in the generated adapters. More elegant, statically verifiable, language-specific implementation options, are mentioned below for each particular case. A well-designed API should use these language features directly, but they might simply not have been available when that part of the API was written:

\paragraph*{Numeric Intervals} Values of a numeric parameter might have to be in an open, closed or mixed interval, e.g. $\left[0, 1\right]$. 
If the type system of the programming language or external linters permit, this requirement can be expressed as a refinement type \cite{refinementTypes}. Then its implementation boils down to putting the stricter parameter type into the adapter's signature.

\paragraph*{Enumerations} A string parameter might only accept a finite set of values. This could be expressed by an enumeration in many languages. Among other benefits, the use of an enumeration improves auto-completion. This solution requires the generation of the additional enumeration type, the replacement of the parameter type in the adapter signature, and the translation of the enumeration values into the values expected by the original API.

\paragraph*{Parameter dependencies} Sometimes, parameter values depend on each other so that not all combinations of parameter values are legal or useful. For instance, a function to serialize an object to JSON could have a parameter that controls whether everything should be dumped into a single line or spread over multiple ones, and another parameter to control indentation. The latter only makes sense in the multiline case. 
Dependent parameters can be avoided by creating an abstract class \code{JSONFormat} with two subclasses \code{Singleline} (with a parameterless constructor) and \code{Multiline} (with a constructor parameter to control indentation). The original two parameters then get replaced by a new one of type \code{JSONFormat}.

\paragraph*{Inference from documentation}
This raises the challenge of inferring the real API of a function, contrasting it to the one expressed in the function declaration, and deriving a suitable transformation. Our solution is based on static analysis of the documentation of each function in an API.
We take advantage of structured documentation formats to break down documentation into pieces that can be analyzed more easily.  

\begin{figure}[h]
    \centering
    \fbox{\includegraphics[width=0.48\textwidth{}]{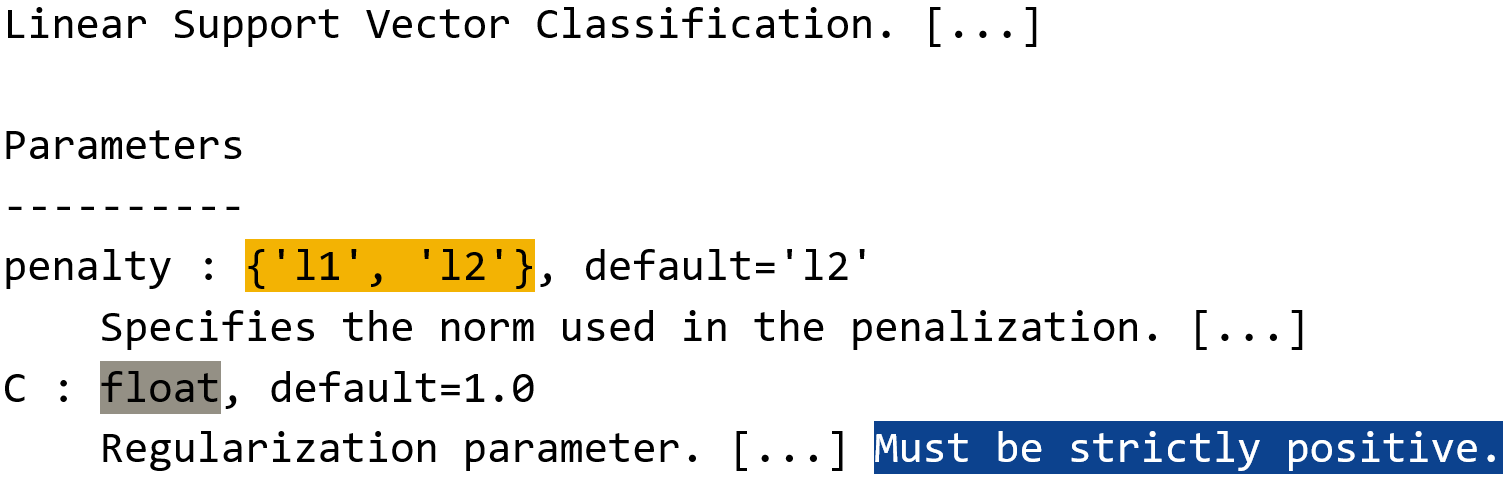}}
    \caption{Example of preconditions hidden in a documentation snippet in Numpydoc format.}
    \label{fig:numpydoc-example}
\end{figure}

\paragraph*{Implementation for Python} 
An added challenge of Python docstrings is that there is no single standard format, but many competing formatting proposals, including reStructuredText \cite{ReStructuredTextPrimerSphinx}, Numpydoc \cite{StyleGuideNumpydoc}, Google's style \cite{GoogleStyleguide}, or Epytext \cite{EpytextMarkupLanguage}. Fig. \ref{fig:numpydoc-example}, for example, shows an excerpt of a docstring for a class in Numpydoc format. It contains a structured description for each parameter of the constructor, with identifiable sections for its name (before the colon), type and default value (after the colon), and a natural-language description (indented in the line below).

\noindent
To implement subsequent steps independently of the docstring format, we first transform them into a common structured representation using the \code{docstring-parser}\footnote{\url{https://github.com/rr-/docstring_parser}}. Afterward, we identify regions that can contain specific preconditions, based on additional textual structure. E.g. the type field (yellow/gray in Fig. \ref{fig:numpydoc-example}) can contain any text, but generally has a relatively fixed format. The set notation \code{\{\textquotesingle{}l1\textquotesingle{}, \textquotesingle{}l2\textquotesingle{}\}} indicates that only these two string literals are legal. This is the clue to suggest the introduction of an enum. The name of the enum is derived from the name of the class/function and parameter, while the names of its instances are derived from the listed values. 

We further analyze the corresponding natural-language description with the rule-based matching of spaCy \cite{Honnibal_spaCy_Industrial-strength_Natural_2020}. For instance, when one of our rules finds the blue text in Fig. \ref{fig:numpydoc-example}, we derive that the parameter \code{C} must be greater than 0.
Once sufficient training data is available, the rule-based matching can be replaced by an ML model to improve generalization.

So far, we only inferred preconditions from docstrings. Analysis results for numeric intervals and enumerations can be improved by including the results of our client code analysis (Sec. \ref{sec:inferring-deletions} and \ref{sec:inferring-optionality-of-parameters}), as suggested by Wang and Zhao \cite{wangPreconditions}.

\section{API Editing}
\label{sec:api-editing}

The automatic inference described in Sec. \ref{sec:inferring-api-transformations} cannot detect all possible API transformations. Moreover, we cannot generally guarantee that a suggested change truly leads to a better API. Ultimately, the redesign of the API is the responsibility of a human. The inferred API transformations can be used as a starting point but require careful \keyword{review} and \keyword{extension} by non-inferrable transformations.

To facilitate the review and extension process (RQ 3), we developed a graphical user interface (GUI) that makes it easy to 
(1) explore the API elements, including their documentation and collected usage data, preventing unnecessary context switches e.g. to the online API documentation,
(2) view the transformations suggested for each element, 
(3) change any transformation and add more, and 
(4) filter the list of API elements to narrow it down to relevant ones. 

\paragraph*{Implementation for Python}
Our proof-of-concept implementation for Python includes a convenient, browser-based GUI, the API Editor\footnote{\url{https://github.com/Safe-DS/API-Editor}}. In the API Editor, we express API transformations as \keyword{annotations} attached to individual API elements. 
Annotations can also capture changes that are not API transformations, e.g. to alter docstrings.

\begin{figure}[h]
    \centering
    \fbox{\includegraphics[width=0.45\textwidth]{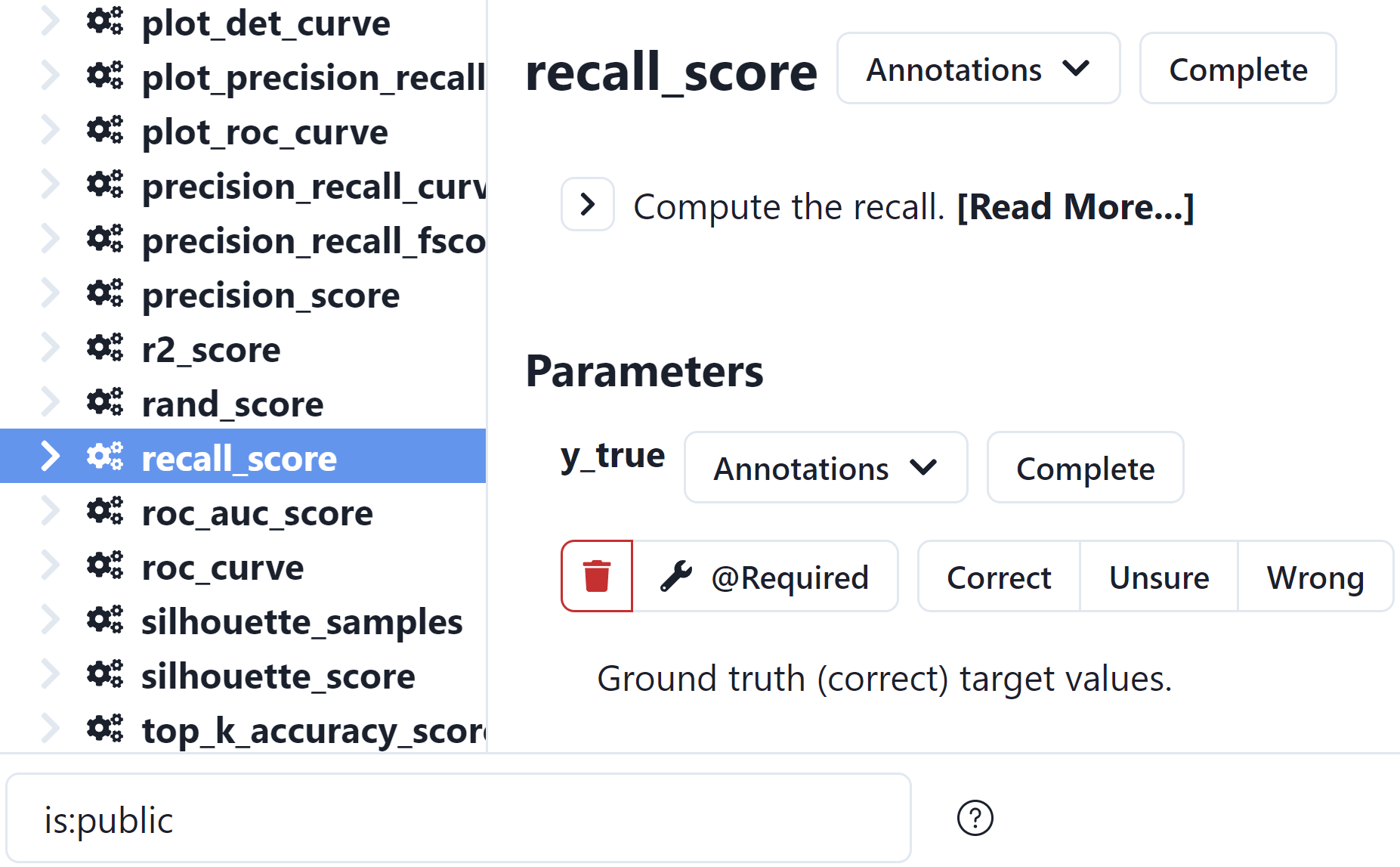}}
    \caption{Excerpt of the GUI of the API Editor. Users can browse through the navigator on the left-hand-side, filter the displayed elements via the input field at the bottom, inspect the annotations of the selected element, and edit annotations. The image shows a selected function and its first parameter.}
    \label{fig:api-editor-screenshot}
\end{figure}

\noindent
Fig. \ref{fig:api-editor-screenshot} shows the main panel of the API Editor\footnote{Some GUI elements were hidden for brevity.}. 
The \keyword{tree view} on the left side shows the different API elements. The selected element (here the function \code{recall\_score}) is marked in blue. Details about the selected element are shown in the \keyword{selection view}, on the right-hand-side. For a function, the selection view displays its documentation, and its parameters. The annotations that are attached to an API element are shown underneath its name, where they can be edited via the button with the wrench symbol or deleted via the trash can symbol. Users can add more annotations by clicking on the ``Annotations'' dropdown next to each element. This opens an annotation-specific editing dialog. For instance, users can specify the new name when editing a \code{Rename} transformation. 

Additionally, the selection view can be used for reviewing the automatically generated annotations or annotations created by a co-developer in a collaborative environment: Individual annotations can be marked as ``Correct'', ``Unsure'', or ``Wrong''. Marking an annotation as ``Unsure'' is a helpful reminder to revisit it later. Marking an annotation as ``Wrong'', instead of deleting it, prevents adding it again and fosters communication among developers.
API elements (e.g. the \code{recall\_score} function and \code{y\_true} parameter shown in Fig. \ref{fig:api-editor-screenshot}), can be marked as ``Complete'', indicating that the review of this element is completed. This disables the annotation dropdown. 

The bottom part of Fig. \ref{fig:api-editor-screenshot} shows the filter box, which is very useful to narrow down a large API. For example, it can be used to show only API elements for which an \code{Add bounds check} transformation has been suggested (Sec. \ref{sec:inferring-enforcement-preconditions}), since these warrant careful manual review.
The GUI has many other features to accelerate the transformation of large APIs, including:

\begin{itemize}
    \item The ability to merge different lists of annotations, which is useful when multiple people edit the same API.
    \item A batch mode to add annotations to all API elements that match the current filter.
    \item Keyboard shortcuts to quickly navigate the tree view or trigger operations.
\end{itemize}
%
The generation of the Python code for the adapters can also be triggered from the GUI (Sec. \ref{sec:adaptoring-approach/adapter-generation}). The created files are packaged into a ZIP file and offered for download.

\section{Dealing with Evolution of the Original Library}
\label{sec:evolution-of-original-library}

The original library may evolve independently of the adapters. As long as its API stays unchanged, the adapters continue to work as expected. If elements are added, we can generate initial adapters for them. If elements get deleted, we can likewise delete the corresponding adapters. However, the API elements for which we already created adapters could also change. Given that the number of breaking changes in an API release is typically limited, they could be processed manually, in the worst case. However, ideally, one would want to avoid tedious specification of API transformations from scratch, since some cannot be inferred and the ones that can be inferred require review (Sec. \ref{sec:inferring-api-transformations}). 
To achieve this (RQ 4), we propose the following approach:

\begin{enumerate}
    \item Record during the API editing process the API transformations applied to the original library to create the current set of adapters. We call this set of changes the \keyword{adapter branch}. 
    \item Automatically derive the \keyword{maintainer branch}, i.e. the API transformations applied to the original library by its maintainers, using a diffing approach like UMLDiff \cite{xing_umldiff_2005}, Diff-CatchUp \cite{xing_api-evolution_2007}, SemDiff \cite{dagenais_recommending_2008,dagenais_semdiff_2009}, AURA \cite{wu_aura_2010}, and HiMa \cite{meng_history-based_2012}.
    
    \item Merge the two sets of API transformations. For this, we can use well-known approaches for automated detection and resolution of conflicts, such as the ones gathered by Mens \cite{mens2002:merging}. Manual intervention for conflicts that cannot be resolved automatically, such as a renaming of the same API element on both branches, can be integrated in the GUI for API editing (Sec. \ref{sec:api-editing}). The user can either keep all changes of the adapter or maintainer branch, or decide on a case-by-case basis.
\end{enumerate}

\paragraph*{Implementation for Python}
The API transformations that were inferred or changed manually can be exported from the GUI as a JSON file (Step 1). The implementation of the other steps of our evolution support approach is ongoing work. 

\section{Evaluation}
\label{sec:evalutation}

We evaluated the quality of the inferred transformations (Sec. \ref{sec:evaluation/quality-inferred-transformations}), the speed of the automated steps of the pipeline (Sec. \ref{sec:evaluation/speed-automated-operations}), and the usability of the GUI (Sec. \ref{sec:evaluation/usability-gui}). Where we refer to \skl{} \cite{sklearn_api, scikit-learn}, we mean version \sklv{}.

Note that the only similar approach to which we could have compared our results, \cite{jugelGeneratingSmartWrapper2010}, provides no transformation inference at all. Usability could not be compared either, due to the lack of a publicly accessible implementation. It is worth noting, however, that the authors themselves mention lack of usability as one of the main shortcomings of their approach.

\subsection{Quality of Inferred Transformations}
\label{sec:evaluation/quality-inferred-transformations}

As shown in Table \ref{tab:implemented-transformations}, we inferred transformations in three categories, using usage-based inference for the first two and text-based inference for the third:
\begin{itemize}
    \item Reduction of API size (\code{Remove}, \code{Replace with constant})
    \item Parameter optionality (\code{Make optional}, \code{Make required})
    \item Enforcement of preconditions (\code{Add bounds check}, \code{Replace with enum})
\end{itemize}

\paragraph*{Usage-based inference}

\begin{figure}[t]
    \centering
    \includegraphics[width=.35\textwidth]{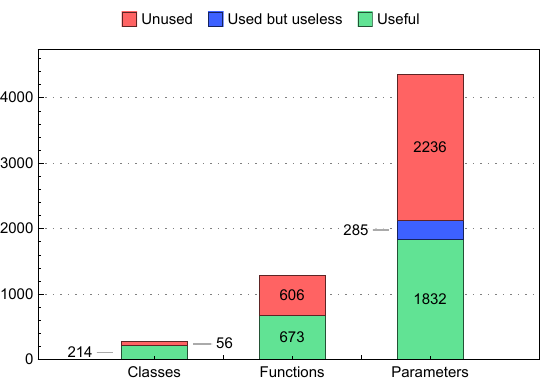}
    \caption{Number of useful, useless and unused elements in the public API of \skl{}. For classes and functions, all used elements are useful.
    } 
    \label{fig:api-size-threshold_1}
\end{figure}

To quantify the achieved size-reduction, we classified API elements of the popular machine learning library \skl{} 
as unused, used but useless, and useful based 
on 41,867 programs\footnote{Mined from \url{https://kaggle.com}.} that use \skl{}.
The results, shown in Fig. \ref{fig:api-size-threshold_1}, indicate that we can reduce the number of classes, functions, and parameters by 20.7\%, 47.4\%, and 57.9\% respectively, if we keep only useful elements.

\paragraph*{Text-based inference} 

We checked the correctness and completeness of the inference of transformations for enforcing preconditions in two separate studies. Initially, we worked with a team of seven students who helped us review the complete API of \skl{}. The results of this evaluation were addressed in a subsequent Bachelor thesis that extended our text-based inference by many cases that were not initially covered. Afterward, we repeated our evaluation, this time on five widely-used libraries, each representing one step of a typical DS workflow. We evaluated the precision on all inferred transformations for the full API of each library. The recall was evaluated on a random sample of 200 functions from each library, since evaluating all five libraries in full would have exceeded our resources. This led to the results reported in Table \ref{tab:text-based-inference/precision} and Table \ref{tab:text-based-inference/recall}. 

\begin{table}[ht]
    \scriptsize
    \centering
    \caption{Precision of Text-Based Inference. The numbers in the table refer to the full library.}
    \begin{tabulary}{.503\textwidth}{|c|c|c|c|c|c|c|c|c|c|} 
        \hline
        \textbf{Library} &
        \multicolumn{3}{c|}{\textbf{Enum}} &
        \multicolumn{3}{c|}{\textbf{Boundary}} &
        \multicolumn{3}{c|}{\textbf{Dependency}} \\
        \hline
               & TP & FP & Prec & TP & FP & Prec & TP & FP & Prec \\
        \hline
        \skl{} & 543 & 45 & 92\% & 234 & 32 & 88\% & 224 & 20 & 92\% \\
        \hline
        Matplotlib & 155 & 18 & 90\% & 14 & 12 & 54\% & 1 & 1 & 50\% \\
        \hline
        Numpy & 95 & 47 & 67\% & 26 & 21 & 55\% & 6 & 11 & 35\% \\
        \hline
        Pandas & 293 & 30 & 91\% & 10 & 14 & 42\% & 21 & 18 & 54\% \\
        \hline
        Seaborn & 16 & 2 & 89\% & 8 & 0 & 100\% & 1 & 0 & 100\% \\
        \hline
    \end{tabulary}
    \label{tab:text-based-inference/precision}
\end{table}

\begin{table}[ht]
    \scriptsize
    \centering
    \caption{Recall of Text-Based Inference, evaluated on a random sample of 200 functions of each library. The numbers in the table refer to these samples.}
    \begin{tabulary}{.503\textwidth}{|c|c|c|c|c|c|c|c|c|c|} 
        \hline
        \textbf{Library} &
        \multicolumn{3}{c|}{\textbf{Enum}} &
        \multicolumn{3}{c|}{\textbf{Boundary}} &
        \multicolumn{3}{c|}{\textbf{Dependency}} \\
        \hline
               & TP & FN & Rec & TP & FN & Rec & TP & FN & Rec \\
        \hline
        \skl{} & 34 & 2 & 94\% & 25 & 7 & 78\% & 21 & 19 & 53\% \\
        \hline
        Matplotlib & 4 & 1 & 80\% & 0 & 0 & - & 0 & 0 & - \\
        \hline
        Numpy & 7 & 6 & 54\% & 0 & 1 & 0\% & 1 & 1 & 50\% \\
        \hline
        Pandas & 14 & 4 & 78\% & 0 & 0 & - & 4 & 5 & 44\% \\
        \hline
        Seaborn & 6 & 0 & 100\% & 4 & 1 & 80\% & 0 & 0 & - \\
        \hline
    \end{tabulary}
    \label{tab:text-based-inference/recall}
\end{table}

We counted a transformation as a true positive (TP) if it did not need changes, as a false positive (FP) if it had to be changed or removed, and as a false negative (FN) if it was missing. 
The results illustrate that the \code{Replace with enum} transformation was relevant to all libraries and that inference worked well, except for Numpy. The FPs and FNs occurred when legal values were only contained in the textual description, rather than the type section of a parameter's documentation. Better matching rules can eliminate this issue.
\code{Add bounds check} and \code{Bundle dependent parameter} (Sec. \ref{sec:inferring-enforcement-preconditions}) were mostly relevant to \skl{}. There, the inference achieved high precision of 88\% and 92\% respectively. The FPs were mostly cases where a boundary or dependency was relativized by the phrases `often', `typically', or `usually', which are currently disregarded by the matching rules. The recall is at 78\% and 53\% respectively. The FNs are caused by slightly different wordings than expected by our text processing rules. These need to be generalized for better recall. 

%




\subsection{Speed of Automated Operations}
\label{sec:evaluation/speed-automated-operations}

On a test PC with Microsoft Windows 11 Pro (Build 22000), an AMD Ryzen 9 3900X processor, Samsung SSD 960 EVO 1TB drive, and 32 GB of main memory, the automated steps performed as follows:
\begin{itemize}
    \item The static analysis of \emph{client code} (Sec. \ref{sec:inferring-deletions} and \ref{sec:inferring-optionality-of-parameters}) needed 28 min 19s to check 92,402 Python programs previously mined from Kaggle, of which 41,867 used \skl{}. This step needs to run only once when new client code is acquired.
    \item The static analysis of \emph{library code} (Sec. \ref{sec:adaptoring-approach/adapter-generation}), including analysis of \emph{docstrings} (Sec. \ref{sec:inferring-enforcement-preconditions}), took 110.3s (mean across 10 runs) to analyze \skl{}. The library contains 422 classes, 2962 global functions or methods, and 8305 parameters. This includes API elements that are marked as internal via naming convention. We include them into our analysis since Python offers no means to hide them, so users could still use them. 
    \item The \emph{adapter generation} (Sec. \ref{sec:adaptoring-approach/adapter-generation}) was completed in 0.86s (mean across 10 runs). It produced 55 Python files with a total of 3,067 lines of code and 20,683 lines of documentation. We determined this with cloc\footnote{\url{https://github.com/AlDanial/cloc}}.
\end{itemize}
We conclude that the efficiency of the automated steps is very good. For each new release of a library, our tool needs below two minutes for analyzing its complete code, inferring sensible improvements, and generating the corresponding adapters. The main limiting factor is the one-time collection  of client programs for our usage analysis. This was slow because of the rate-limited API of Kaggle. Other interesting sources, e.g. GitHub, have similar intentional limitations. 

\subsection{Usability of GUI}
\label{sec:evaluation/usability-gui}



We evaluated the usability of the API Editor in a study with 5 bachelor students, who 
were familiar with Python, refactoring and design patterns but had no prior knowledge of the API Editor.
Since no public prototype of the only closely related approach \cite{jugelGeneratingSmartWrapper2010} is available, we compared the automatic adapter generation via our API Editor (version 1.64.1) to what we consider the best state-of-the-art tool for (partly) automated creation of adapters: Use of GitHub copilot\footnote{\url{https://github.com/features/copilot}} (version 1.1.29.1869) in PyCharm (version 222.3345.131). GitHub copilot excels at suggesting code that obeys strict patterns, like adapters. PyCharm offers the most comprehensive support for Python programming, in our opinion.

Our study compared these two approaches by creating adapters for \mplt{} v3.5.3 \cite{Hunter:2007}. This is a popular API for plotting, a domain that can be understood intuitively.

%
%
%
To limit fatigue, the study consisted of four main blocks, each constrained to 30 minutes, with two 15-minute breaks in-between. 
The session was recorded with permission and participants were asked to think aloud \cite{ericssonProtocolAnalysisVerbal1984, borenThinkingAloudReconciling2000, chartersUseThinkaloudMethods2003, jaspersThinkAloudMethod2004, youngDirectSourceValue, ecclesThinkAloudMethod2017} while working on the tasks.

In the first block, 
we introduced 10 general usability guidelines based on \cite{myersImprovingAPIUsability2016}, to give participants an idea about good API design without biasing them towards the changes that can be made with the API Editor. 

In the second block, 
the participants were asked to find usability issues in an area of \mplt{} of their choice, based on its API documentation\footnote{\url{https://matplotlib.org/stable/api/index.html}} and the source code linked from there. 
Participants could creatively come up with their own usability and design flaws. This was again intended to prevent potential bias towards our tool. The participants wrote down each issue they found, including its precise location in the API and a possible resolution.

In the third block, we briefly explained the transformations supported by the API Editor and provided as reference a specification similar to Table \ref{tab:implemented-transformations}. The participants then used the API Editor to resolve the issues they had found in the previous step. Once they were done, they filled out a standard \keyword{System Usability Scale} (SUS) questionnaire (\cite{brookeSUSQuickDirty, bangorEmpiricalEvaluationSystem2008, mclellanEffectExperienceSystem2011, lewisSystemUsabilityScale2018}) about their experience of using the API Editor on the \mplt{} API. 

In the final block, we reminded participants of the adapter design pattern and explained its implementation in Python for global functions and instance methods of classes. 
The participants were again provided with a reference. 
Then, they were asked to implement adapters with GitHub Copilot in PyCharm, for the changes that they had previously performed with the API Editor. 
Finally, the participants filled out a SUS questionnaire about the usability of GitHub Copilot and PyCharm for improving the \mplt{} API.

The API Editor received scores between 82.5 and 90 with a median of 90. Everything above 85.5 corresponds to excellent usability on the SUS scale. The scores regarding the use of PyCharm for this task varied between 22.5 and 70 with a median of 55. The lowest score, however, was given by a person without prior experience with PyCharm, so we regard it as an outlier. This lifts the median to 58.75. Still, the highest score, 70, is just slightly higher than the SUS `above average' mark of 68. Thus, the API Editor provided a significantly improved  user experience.

%




We also counted how often participants succeeded or failed in implementing fixes for the design flaws that they had found in \mplt{}. Then we compared the results for the \apieditor{} and GitHub Copilot.
All renames (41), removals (12), and additions of bounds checks (2) were performed correctly in the \apieditor{}. The attempts to combine two functions (3) or split a function into two (3) could not be done in the \apieditor{}, since they are not generally adaptorable.
%
%
No participant managed to implement all the adapters correctly with PyCharm, though. 14 of 41 attempts to rename a function using an adapter were implemented wrongly (often, multiple issues affected the same adapter). Common mistakes included: 
\begin{itemize}
    \item Incorrect parameter list of an adapter (9 cases).
    \item Passing a keyword-only argument by position (9 cases).
    \item Passing the original default value in the call to the original library, instead of the value of the corresponding parameter of the adapter (3 cases).
\end{itemize}
Experienced Python programmers might not make these mistakes. Still, the evaluation clearly showed that the \apieditor{} is valuable, especially for novices, to prevent errors caused by uncommon language features.

\subsection{Threats to Validity}
\label{sec:evaluation/threats-to-validity}

\paragraph*{Participants of usability study} All participants of the usability study (Sec. \ref{sec:evaluation/usability-gui}) were bachelor students. More experienced developers would be able to implement the adapters with fewer bugs, which we could also see in our small target group. Likewise, better training would improve the performance. But even then, bugs would be inevitable when creating adapters for all elements of a large library. Moreover, training developers or employing more senior ones would increase cost. Finally, adapter creation is a rather boring task that would negatively impact the happiness of developers, particularly senior ones.

\paragraph*{Design of usability study} Instead of finding usability issues by writing own code that \emph{uses} \mplt{}, participants of the usability study (Sec. \ref{sec:evaluation/usability-gui}) were asked to find flaws just by looking at the documentation and library code, due to time constraints. This way, participants might have only detected simple usability problems and missed more in-depth issues \cite{grillMethodsAPIUsability2012} for which no annotation exists yet in the API Editor. Moreover, even though all tasks were constrained to 30 minutes with breaks in between, fatigue might have had a negative impact on the SUS scores of the second experiment (manual adapter creation via PyCharm and GitHub Copilot). However, we do not expect the large gap between the SUS scores of the two different approaches to vanish completely.

\paragraph*{Applicability to other libraries} 
Our findings regarding the usability of the GUI (Sec. \ref{sec:evaluation/usability-gui}) can be generalized to any library, since the GUI works the same for them all. Likewise, we expect the performance of the automated parts of the pipeline (Sec. \ref{sec:evaluation/speed-automated-operations}) to be sufficient, since \skl{} is already fairly large. The quality of inferred transformations (Sec. \ref{sec:evaluation/quality-inferred-transformations}) depends on the existence of high-quality docstrings in the library code and the availability of large amounts of client code, though.

\section{Related Work}
\label{sec:related-work}

\paragraph*{API complexity} Removing unused and useless API elements follows the API usability principle of ``aesthetic and minimalist design'' \cite{myersImprovingAPIUsability2016}. The empirical study \cite{schellerAutomatedMeasurementAPI2015} suggests that having more API elements inside the same namespace reduces the success rate of finding the ones needed for a given task. Stylos and Myers found that this effect could be mitigated by proper naming and particularly distinct name prefixes to make full use of auto-completion \cite{stylosImplicationsMethodPlacement2008}. Unnecessary API complexity was also the second most common issue identified in the Contextual Interaction Framework during a heuristic evaluation \cite{grillMethodsAPIUsability2012}. Lämmel et al. \cite{lammelLargescaleASTbasedAPIusage2011}
analyzed usage patterns of Java APIs that are useful for API migration. 

Several guidelines exist regarding the number of parameters a function should have: Martin \cite[p. 288]{clean_code} suggests avoiding more than three parameters. McConnell's guideline \cite[p. 178]{code_complete} is to use at most seven parameters, based on psychological research. Stylos and Clarke \cite{stylosUsabilityImplicationsRequiring2007} even found parameterless constructors with subsequent setters to offer better usability than constructors with parameters.
Complexity is also a recurring theme in usability metrics for APIs \cite{desouzaAutomaticEvaluationAPI2009, ramaStructuralMeasuresAPI2015, schellerAutomatedMeasurementAPI2015}.

Debloating \cite{xin-debloating-2020, xin-debloating-2023} offers a way to produce a reduced program that still works correctly for \textit{existing clients} on a given set of inputs. Adaptoring, meanwhile, creates a new, reduced API, designed to improve learnability and reduce the risk of misuse by \textit{new clients}. It does not reduce the memory footprint of the program but increases it, adding the adapter code to the code of the original library, which must still be installed. 

\paragraph*{API usability} General guidelines for high-quality code like \cite{clean_code, code_complete, refactoring} apply especially to public APIs. Additionally, Bloch \cite{blochHowDesignGood2006a} as well as Myers and Stylos \cite{myersImprovingAPIUsability2016} provide specific guidance for API design. 
Many more studies exist about this topic: 
Hou and Li \cite{houObstaclesUsingFrameworks2011} investigated discussions from the Java Swing forum to find common learning hurdles. Grill et al. \cite{grillMethodsAPIUsability2012} evaluated the usability of an API using various methods from the area of human computer interaction. Scheller and Kuhn \cite{schellerInfluencingFactorsUsability2012} compared two API designs in a study to determine their usability when instantiating classes or calling methods. 
Murphy et al. \cite{murphyAPIDesignersField2018} aimed to better understand the process behind API design through developer interviews.

\paragraph*{Code smells} Given the similarity of adaptorable API transformations and refactorings, inference of recommendable API transformations is akin to the detection of code smells \cite{refactoring}, which indicate places where refactorings should be applied. For example, a field with a \code{Mysterious Name} is an opportunity to apply the \code{Rename Field} refactoring.
Lacerda et al. \cite{lacerdaCodeSmellsRefactoring2020} reviewed literature about the definition of code smells, means to detect them, and their correspondence to refactoring. They also found that code smells and refactorings are strongly linked to code quality dimensions, e.g. understandability. Santos et al. \cite{santosSystematicReviewCode2018}, however, concluded that no strong correlation between code smells and code quality exists, and that developers rarely agree with detected code smells. Yamashita and Moonen \cite{yamashitaDevelopersCareCode2013} found that developers with higher knowledge about code smells also found them more important. Fontana et al. \cite{arcellifontanaAutomaticDetectionBad2012}, Al Dallal \cite{aldallalIdentifyingRefactoringOpportunities2015}, Rasool and Arshad \cite{rasoolReviewCodeSmell2015}, and Di Nucci et al. \cite{dinucciDetectingCodeSmells2018} surveyed approaches to detect code smells. This field of research is too wide to fully do it justice here.


\section{Limitations and Future Work}
\label{sec:limitations-future-work}

\paragraph*{Static analysis of Python code} We opted to use purely \emph{static} analysis of the library and client code, so users would not accidentally execute arbitrary code pulled from the internet on their machine.

Because of this, our prototype implementation cannot accurately capture the entire API of every Python library.
For instance, Python's decorators (PEP 318 \cite{smithPEP318Decorators}) can be used to alter the signature of a function 
at runtime. This can be overcome by using the \code{inspect} module \cite{InspectInspectLive}, which offers introspection at runtime and can capture dynamic signature modifications. 
The \code{inspect} module can be used together with library code without additional security risks, since library code must run on the user's machine anyway when they use the generated adapters. Since library developers are generally careful not to execute slow code when a module gets imported, the performance of the analysis should still suffice.

Another challenge pose dynamic imports \cite{ImportlibImplementationImport}, which make it statically impossible to determine all call targets in client code. This limits the quality of our usage analysis and can only be overcome by fully dynamic analysis. However, measures must be taken for sandboxing of analyzed client code. Also, the greatly increased runtime of the analysis must be considered, e.g. if client code is interspersed with the training of machine learning models.

\paragraph*{Inference of API transformations}
The usage-based analysis (Sec. \ref{sec:inferring-deletions} and \ref{sec:inferring-optionality-of-parameters}) needs representative client code. The API designer is responsible for collecting it from general-purpose sources, such as GitHub, or domain-specific sources like Kaggle. They must also set the threshold and significance level appropriately, depending on the available client code.
However, unless we analyze all client code, which is impossible for open-source libraries, the absence of usages of a function in the analyzed client code does not imply that the function is not used in general. As Hyrum's Law \cite{wrightHyrumLaw} suggests, with enough users, all API elements will eventually be used. 
Moreover,
we cannot always derive the target of a call, as explained above. Therefore, we might incorrectly mark a function as unused that is used in the analyzed client code.

Additionally, we don't consider \emph{when} a function was added to a library: A more recent addition is likely to be used less than a function that has been part of the library for years. We can similarly discuss the correctness of other automated suggestions. The main takeaway here is that, ultimately, the user of the tool is responsible for the redesign of the API. The automated suggestions are meant to accelerate the redesign process, but must be reviewed carefully. 


\paragraph*{Applicability to other languages} Our prototype implementation can only handle Python libraries. This, however, is not a limitation of the adaptoring approach. On the contrary, it would be easier to handle a language like Java, whose enforced static typing eases  thorough static analysis of library and client code, and whose standardized, rich documentation format, greatly simplifies text-based analysis. 

\paragraph*{Additional API transformations} We only implemented a subset of the adaptorable API transformations (Sec. \ref{sec:adaptoring-approach/adaptorable-api-transformations}) to test the different elements of our approach, including the analysis of library code, the contained docstrings, and client code. Additional transformations can be added later. But, as with refactorings, a transformation must be needed often enough to justify the effort to automate it. Thus, there will always be cases where manual changes are required. 
\paragraph*{Manual changes to adapters} Currently, we don't offer a way to edit the generated adapters and preserve the edits when  adapters are re-generated. To support this, manually adjusted adapters must be detected and no longer overridden. This entails that the previous adapters become an input to the adapter generation process.

\section{Conclusion}
\label{sec:conclusion}

Adaptoring can provide a new API for an existing library to fix learnability and usability issues or support different client profiles.
The new API is realized as a wrapper library that calls the original library. Because the original library remains untouched, adaptoring \textit{avoids} breaking existing clients and the need to keep deprecated API elements. Because no code is duplicated, adaptoring can leverage bug fixes or performance improvements to the original library.
Most API-transforming refactorings can be realized by adaptoring (Sec. \ref{sec:adaptoring-approach/adaptorable-api-transformations}, RQ 1).

To make adaptoring viable for large libraries, many tasks are automated: Given a list of API transformations, adapter code gets generated automatically (Sec. \ref{sec:adaptoring-approach/adapter-generation}). Unlike in \cite{jugelGeneratingSmartWrapper2010}, a subset of API transformations can be inferred automatically (RQ 2): Combined usage and documentation analysis detects opportunities for deletion of useless API elements, 
making parameters optional or mandatory, 
and enforcement of various preconditions.
The inferred transformations can be reviewed and edited 
by a human in a novel graphical user interface, 
(Sec. \ref{sec:api-editing}, RQ 3), 
which received an excellent median system usability score of 90 (Sec. \ref{sec:evaluation/usability-gui}).

We implemented our approach for Python and applied it to popular libraries for data science. Our text-based inference of preconditions achieves high precision (88 to 92\%) with varying recall (53 to 94\%) on \skl{}. Improving recall is mainly an issue of adding further text processing rules to cover yet missed cases (Sec. \ref{sec:evaluation/quality-inferred-transformations}). 
All automated steps together took less than 31 minutes on the \skl{} library and 41,867 client programs (Sec. \ref{sec:evaluation/speed-automated-operations}). 

We also investigated how to keep the generated adapters updated automatically in case the API of the original library changes (RQ 4) by treating the API transformations applied by the maintainers of the original library and by the adaptorers as two separate branches that need to be merged (Sec. \ref{sec:evolution-of-original-library}). Implementation of this step is ongoing.

Overall, adaptoring allows making an improved API available to the user community with little effort, either as the finished result or as an intermediate step to gather feedback. If necessary, the API can then be refined further in fast iterations.

\section*{Acknowledgments}
Nils Vollroth implemented the text-based inference of API transformations and performed the evaluation summarized in Tables \ref{tab:text-based-inference/precision} and \ref{tab:text-based-inference/recall}. The current state of the project would not have been possible without the numerous enthusiastic contributors to the repository. The insightful comments of the anonymous reviewers helped improve the final paper.

\bibliographystyle{IEEEtran}
\bibliography{IEEEabrv,references}

\end{document}